\documentclass[useAMS,usenatbib]{mn2e}

\usepackage{epsf}
\usepackage{graphicx}
\usepackage{color}
\usepackage{natbib}

\title[The impact of dust on the scaling properties of galaxy
clusters]{The impact of dust on the scaling properties of galaxy
clusters}

\author[A. da Silva et al.]{
A. da Silva$^{1,2}$\thanks{E-mail: asilva@astro.up.pt}; 
A. Catalano$^{3}$, 
L. Montier$^{4}$, 
E. Pointecouteau$^{4}$, 
J. Lanoux$^{4}$, 
M. Giard$^{4}$\\ 
$^{1}$Centro de Astrofisica da Universidade do Porto, Rua das Estrelas, 4150-762 Porto, Portugal \\
$^{2}$Institut d'Astrophysique Spatiale, Bat 121, Universite Paris Sud, 91405 Orsay, France\\
$^{3}$Observatoire de Paris - LERMA, 61 avenue de l'Observatoire, 75014 Paris, France\\
$^{4}$CESR, CNRS - Universit\'e de Toulouse, BP 44346, F 31028, Toulouse cedex 04, France
}
\begin{document}
%
\def\etal{et  al.\ }
\def\araa{{Ann.\ Rev.\ Astron.\ Ap.}}
\def\aplet{{Ap.\ Letters}}
\def\aj{{Astron.\ J.}}
\def\apj{ApJ}
\def\apjl{{ApJ\ (Lett.)}}
\def\apjs{{ApJ\ Suppl.}}
\def\aas{{Astron.\ Astrophys.\ Suppl.}}
\def\aa{{A\&A}}
\def\aap{{A\&A}}
\def\mnras{{MNRAS}}
\def\nat{{Nature}}
\def\pasa{{Proc.\ Astr.\ Soc.\ Aust.}}
\def\pasp{{P.\ A.\ S.\ P.}}
\def\pasj{{PASJ}}
\def\pre{{Preprint}}
\def\sovlet{{Sov. Astron. Lett.}}
\def\adspr{{Adv. Space. Res.}}
\def\expas{{Experimental Astron.}}
\def\ssr{{Space Sci. Rev.}}
\def\apss{{Astrophys. and Space Sci.}}
\def\inpress{in press.}
\def\souspresse{sous presse.}
\def\inprep{in preparation.}
\def\enprep{en pr\'eparation.}
\def\submit{submitted.}
\def\soumis{soumis.}
\def\aph{{Astro-ph}}
\def\astroph{{Astro-ph}}


\pagerange{\pageref{firstpage}--\pageref{lastpage}} \pubyear{2002}

\maketitle

\label{firstpage}

\begin{abstract}
We investigate the effect of dust on the scaling properties of galaxy
clusters based on hydrodynamic $N$-body simulations of structure
formation. We have simulated five dust models plus a radiative cooling
and adiabatic models using the same initial conditions for all
runs. The numerical implementation of dust was based on the analytical
computations of \citet{montier04}.  We set up dust simulations to
cover different combinations of dust parameters that put in evidence
the effects of size and abundance of dust grains. Comparing our
radiative {\it plus} dust cooling runs to a purely radiative cooling
simulation we find that dust has an impact on cluster scaling
relations. It mainly affects the normalisation of the scalings (and
their evolution), whereas it introduces no significant differences on
their slopes.  The strength of the effect critically depends on the
dust abundance and grain size parameters as well as on the cluster
scaling.  Indeed, cooling due to dust is effective at the cluster
regime and has a stronger effect on the ``baryon driven'' statistical
properties of clusters such as $L_{\rm X}-M$, $Y- M$, $S-M$ scaling
relations. Major differences, relative to the radiative cooling model,
are as high as 25\% for the $L_{\rm X}-M$ normalisation, and about
10\% for the $Y- M$ and $S-M$ normalisations at redshift zero.  On the
other hand, we find that dust has almost no impact on the ``dark
matter driven'' $T_{\rm mw}-M$ scaling relation.  The effects are
found to be dependent in equal parts on both dust abundances and grain
sizes distributions for the scalings investigated in this
paper. Higher dust abundances and smaller grain sizes cause larger
departures from the radiative cooling (i.e. with no dust) model.
\end{abstract}


\section{Introduction}

From the first stages of star and galaxy formation, non-gravitational
processes drive together with gravitation the formation and the
evolution of structures. The complex physics they involve rule the
baryonic component within clusters of galaxies, and in a more general
context within the intergalactic medium (IGM hereafter -- see the
review by \citet{voit05} and references therein). The study of these
processes is the key to our understanding of the formation and the
evolution of large-scale structure of the Universe. Indeed,
understanding how their heating and cooling abilities affect the
thermodynamics of the IGM at large scales and high redshifts, and thus
that of the intra-cluster medium (hereafter ICM) once the gas get
accreted onto massive halos is a major question still to be
answered. The continuous accretion and the merger events through which
a halo assembled lead to a constant interaction of the IGM gas with
the evolving galactic component. Within denser environments, like
clusters, feedback provided by AGN balances the gas cooling (see for
instance \citet{cattaneo07,conroy07}, and \citet{mcnamara07} for a
review). Also, from high redshift, the rate of supernovae drives the
strength of the galactic winds and thus the amount of material that
ends ejected within the IGM and the ICM (see
\citet{loewenstein06}). These ejecta are then mixed in the environment
by the action of the surrounding gravitational potential and the
dynamics of cluster galaxies within.

Since long, X-ray observations have shown the abundant presence of
heavy elements within the ICM (see for instance review works by
\citep{sarazin88,arnaud05}). Physical processes like ram-pressure
stripping, AGN interaction with the ICM, galaxy-galaxy interaction or
mergers are scrutinized within analytical models and numerical
simulations in order to explain the presence of metals (see for
instance works by \citep{kapferer06, domainko06,moll07}). Moreover, it
is obvious that the process of tearing of material from galaxies leads
not only to the enrichment of the ICM/IGM in metal, but in gas, stars
and dust as well. 
Recent work on numerical simulations
\citep{murante04,murante07,conroy07} have stressed the role of
hierarchical buiding of structures in enriching the ICM with stars in
a consistent way with the observed amount of ICM globular clusters,
and ICM light. Indeed, the overall light coming from stars in between
cluster galaxies represent an important fraction of the total cluster
light: for instance \citep{krick07} measured 6 to 22\% from a sample
of 10 clusters.  The effect of a diffuse dust component within the
IGM/ICM, and its effect is less known. A few observational studies
with the ISO and the Spitzer satellites have tried without frank
success to detect the signature of such a component \citep{stickel98,
stickel02, bai06, bai07}. More successfully, \citep{montier05} have
obtained a statistical detection, via a stacking analysis, of the
overall infrared emission coming from clusters of galaxies. However,
they were not able to disentangle the IR signal from dusty cluster
galaxies from a possible ICM dust component. On the other hand, from a
theoretical point of view a few works have looked at the effect of
dust on the ICM \citep{popescu00} or in conjunction with the
enrichment of the ICM in metals \citep{aguirre01}. However, the effect
of dust on a ICM/IGM-type thermalized plasma has been formalized by
\citep{montier04}. These authors have computed the cooling function of
dust taking into account the energetic budget for dust. They have
shown the ability of dust to be a non negligible cooling/heating
vector depending on the physical properties of the
environment.

Dust thus comes, within the ICM/IGM, as an added source of
non-gravitational physics that can potentially influence the formation
and the evolution of large scale structure in a significant
way. Indeed, since redshift of $z\simeq 2-5$ during which the star
formation activity reached its maximum in the cosmic history, large
amounts of dust has been produced and thus ejected out of the galaxies
due to violent galactic winds into the IGM \citep{springel03}. As this
material is then accreted by the forming halos, one can wonder about
the impact produced by dust on the overall properties of clusters of
galaxies once assembled and thermalized. In a hierarchical Universe,
the population of clusters is self-similar, thus is expected to
present well defined structural and scaling properties. However, to
date, it is common knowledge that the observed properties deviate form
the prediction by a purely gravitational model (see \citep{voit05,
arnaud05} for review works). It is thus important to address the issue
of the impact of dust on the statistical properties of structures such
as clusters of galaxies, the same way it is done for AGNs, supernovae,
stripping or mergers.

In order to tackle this question, we have put into place the first
N-body numerical simulations of hierarchical structure formation
implementing the cooling effect of dust according to the dust
nature and abundance. In this paper, we present the first results of
this work focusing at the scale of galaxy clusters, and more
specifically on their scaling properties. The paper is organized as
follows: we start by presenting the physical dust model and
how it is implemented in the numerical simulation code. In Sec
.~\ref{sec:sinum}, we describe the numerical simulations and the
various runs (i.e. model) that have been tested. From these simulations
our analysis concerns the galaxy cluster scale, and focus on the
impact of the presence of dust on the scaling relation of clusters. In
Sec.~\ref{sec:scal}, we present our results on the $M-T$, the $S-T$,
the $Y-T$ and the $L_X-T$ relations. The derived results are presented
in Sec.~\ref{sec:res} and discussed in Sec.\ref{sec:dis}.

\section{The dust model}
\label{sec:dustmod}
In our numerical simulations the implementation of the physical effect
of dust grains is based on the computation by \citet{montier04} of the
dust heating/cooling function. In this work, we decided to limit our
implementation to the dust cooling effect only. Indeed the goal of
this paper is to study the effect of dust at the galaxy cluster
scales. The heating by dust grains is mainly effective at low
temperatures (i.e $T_e<10^5$~K) and is a localised effect strongly
dependent of the UV radiation field. Our numerical simulations (see
Sec.~\ref{sec:sinum} and \ref{sec:limit}) do not directly implement
this level of physics.

Dust grains in a thermal plasma with $10^6 < T < 10^9$~K are destroyed
by thermal sputtering, which efficiency was quantified by \citet[see
their Eq.~44]{draine79}. The sputtering time depends on the column
density and on the grain size. For grain sizes ranging form $0.001
\mu$m to $0.5 \mu$m, and an optically thin plasma ($n\sim
10^{-3}$~atom/cm$^2$), the dust lifetime spawns from $10^6$~yr for
small grains up to $10^9$~yr for big grains. This lifetimes are
therefore large enough for the cooling by dust in the IGM/ICM to be
considered. Evidently, it is also strongly linked to the injection
rate of dust, thus to the physical mechanism that can bring and spread
dust in the IGM/ICM.

Our implementation of the dust cooling power is based on the model by
\citep{montier04}. We recall bellow the main aspects of this model and
describe the practical implementation within the $N$-body simulations.

\subsection{The dust cooling function}
Dust grains within a thermal gas such as the ICM or the IGM can either
be a heating or a cooling vector depending on the physical state of
the surrounding gas and on the radiative environment. Heating can
occur via the photo-electric effect if the stellar radiation field
(stars and/or QSOs) is strong enough (\citet{weingartner06} and
references therein). Indeed, the binding
energies of electrons in dust grains are small, thus allowing
electrons to be more easily photo-detached than in the case of a free
atom or a molecule.  On the other hand, the cooling by dust occurs
through re-radiation in the IR of the collisional energy deposited on
grains by impinging free electrons of the ICM/IGM \footnote{In the
galactic medium the cooling occurs through re-radiation of the power
absorbed in the UV and visible range.}.

\citet{montier04} have computed the balance of
the heating and cooling by dust with respect to the dust abundance:
cooling by dust dominates at high temperatures in the hot IGM of
virialized structures (i.e clusters of galaxies), and heating by dust
dominates in low temperature plasma under high radiation fluxes such
as in the proximity of quasars. The details, of course, depend on the
local physical parameters such as the grain size and the gas density.

Assuming local thermal equilibrium for the dust, the overall balance
between heating and cooling in dust grains can be written as
follows:
\begin{equation}
\Lambda^g(a, T_d) = H^g_{coll}(a, T_e, n_e),
\label{eq:theq}
\end{equation}
with $H_{coll}$ being the collisional heating function of the grain
and $\Lambda$ the cooling function due to thermal radiation of
dust. $a$ is the grain size, $T_e$ and $n_e$ are respectively the
electronic temperature and density of the medium and $T_d$ is the dust
grain temperature.

The heating of the dust grain was taken from \citet{dwek81} and can be
expressed in a general way as:
\begin{equation}
H^g_{coll}(a, T_e, n_e) \propto n_e\, a^\alpha\, T_e^\beta
\label{eq:theq2}
\end{equation}
where the values of $\alpha$ and $\beta$ are dependent of the value of
the ratio $a^{2/3}/T_e$.

The relevant dust parameters affecting the cooling function are the
grain size and the metallicity. Indeed, the smaller the grains and the
higher the metallicity, the higher is the cooling power of the
dust. Thus the total cooling function due to a population of dust
grains can be expressed as a function of these two parameters as:
\begin{equation}
\Lambda(a, T_d) = \int\int\int \Lambda^g(a,T_d)\frac{\textrm{d}
N(a,Z,V)}{\textrm{d}V \textrm{d}a \textrm{d}Z} \textrm{d}V \textrm{d}a
\textrm{d}Z
\label{eq:theq3}
\end{equation}
where $\textrm{d} N(a,Z,V)/\textrm{d}V \textrm{d}a \textrm{d}Z$ is the
differential number of dust grains per size, metallicity and volume
element.

Cooling by dust happens to increase with the square root of the gas
density, whereas the heating by dust is proportional to the
density. As stressed by \citet{montier04} the cooling by dust is more
efficient within the temperature range of $10^6<T<10^8$~K (i.e
$0.1<kT<10$~keV), which is typically the IGM and ICM thermal
conditions.

We redirect the reader to \citet{montier04} for a full description of
the dust model, and a comprehensive physical analysis of the effect of
dust in a optically thin plasma.

\subsection{The dust abundance}
The abundance of dust is a key ingredient to properly weight in our
implementation. Observations indicate that dust represents only a
tiny fraction of the baryonic matter: $M_{dust}/M_{gas} \approx 0.01$
in our Milky Way \citep{dwek90}, and this is possibly lower by a
factor 100 to 1000 in the ICM: $M_{dust}/M_{gas} = 10^{-5}-10^{-4}$
\citep{popescu00,aguirre01}. We defined the abundance of dust as the
ratio of the dust mass with respect to the gas mass:
\begin{equation}
Z_d  = \frac{M_{dust}}{M_{gas}} = f_d\; \frac{Z}{Z_\odot}\; Z_{d\, \odot}
\label{eq:zd}
\end{equation}
where $Z$ is the metallicity in units of solar metallicity, $Z_{d\,
\odot}=0.0075$ is the solar dust abundance, i.e the dust-to-gas mass
ratio in the solar vincinity \citep{dwek90}, and $f_d$ is the
abundance of dust in the ICM in units of solar dust abundance.

Dust enrichment occurs via the feedback of galaxy formation and
evolution in the ICM through interaction, stripping, mergers, galactic
winds and AGNs outburst. At all redshifts, it is linked to the SFR
which drives the production of dust in cluster galaxies. However, in
our hydrodynamic simulations (see Sect.~\ref{sec:sinum}) the SFR is
not physically modeled, but it is inferred by the cooling state of the
gas particles within the simulations: gas particles below a given
threshold of temperature and above a given threshold of density are
considered as colisionless matter, forming stars and galaxies (see
Sec.~\ref{sec:sinum}). In order to tackle this problem, we choose to
directly link the dust abundance to the metal abundance using
Eq.~(\ref{eq:zd}). Therefore, the dust distribution in our simulations
mimics the metal distribution.

\subsection{Implementation in the $N$-body simulations}
\label{sec:impmod}

From the equations presented in the previous sections, we computed
the dust cooling function according to the embedding medium
temperature and (global) metallicity.
In simulations, once the metallicity and temperature
are known, $a$ and $f_d$ are the only two parameters driving the dust
cooling rate (i.e $\Lambda(a,Z)=\Lambda(a,f_d)$). In the top panel of
Fig.~\ref{fig:lamdust} we present dust cooling rates (red lines) for
$f_d=0.1$ and $a=10^{-3}$~$\mu$m (model D1, see below) at different
values of metallicity. The blue and black lines are
the radiative cooling rates from \citet{sutherland:1993} and the total
(i.e radiative plus dust cooling) rate, respectively.

\begin{figure}
\begin{center}
\epsfxsize=8.50cm \epsfbox{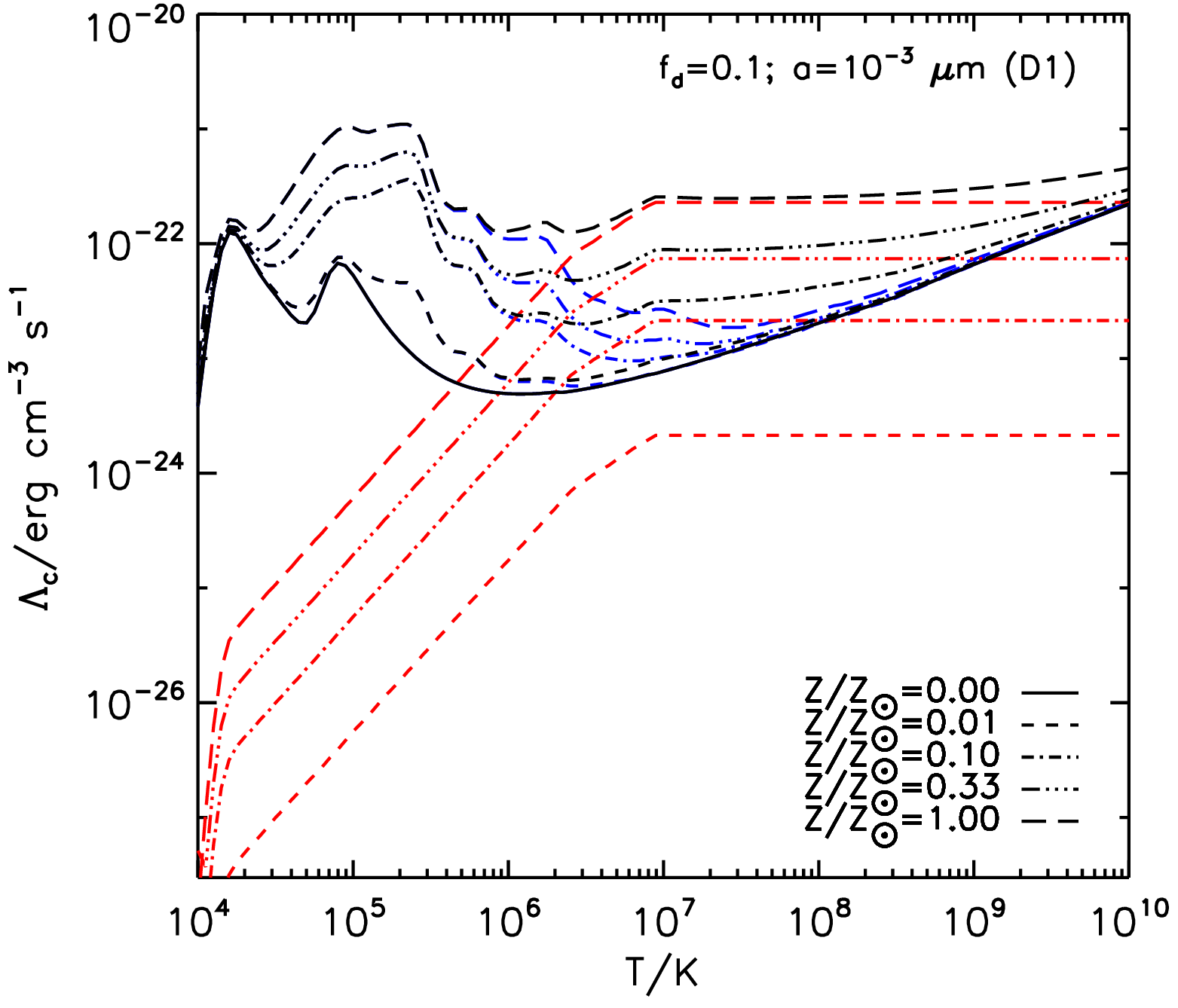}
\epsfxsize=8.50cm \epsfbox{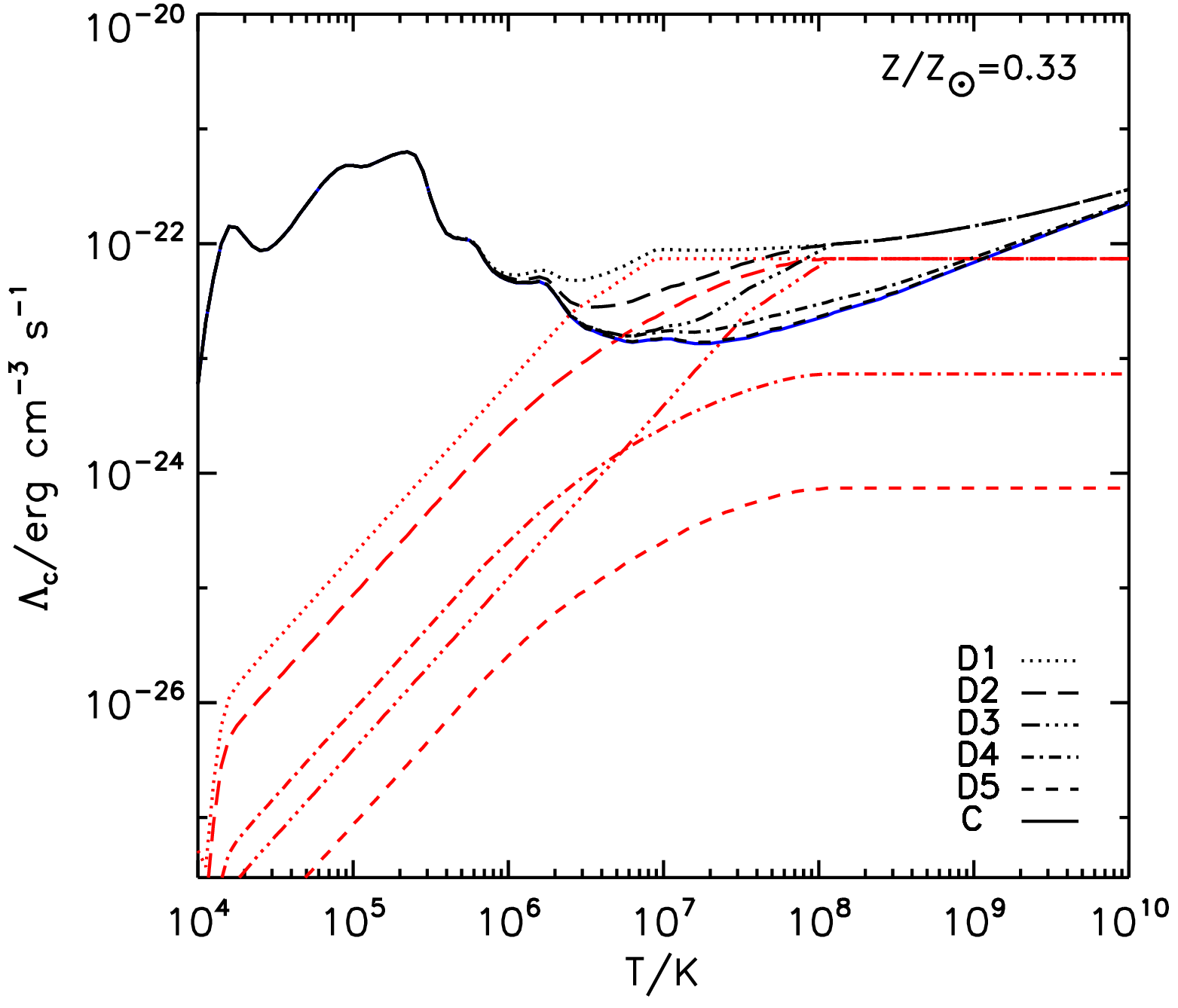}
\caption{\label{fig:lamdust} Cooling functions implemented in the
numerical simulations. Top panel shows the dependence of dust model D1
($f_d=0.1$ and $a=10^{-3}$~$\mu$m) with metallicity
(and temperature) whereas the bottom panel shows different dust models
at the same metallicity $Z/Z_\odot=0.33$ (see
text). Black, blue and red curves are the total cooling functions,
radiative cooling of the gas from \citet{sutherland:1993} and dust
cooling functions, respectively.
}
\end{center}
\end{figure}

Together with an adiabatic run (i.e model A)
and a ``standard'' radiative run (model C -- see Sect.~\ref{sec:sinum}
for further details), we ran a total of five runs implementing various
population of grains (i.e named D1 to D5) characterized by their sized
and dust-to-metal mass ratio:\\[-1.8em]
\begin{itemize} 
\item We tested three types of sizes: two fixed grain sizes with
  $a=10^{-3}$~$\mu$m and $a=0.5$~$\mu$m), respectively labeled {\it
  small} and {\it big}. The third assumes for the IGM dust grains a
  distribution in sizes as defined by \citet{mathis77} for the
  galactic dust: $N(a)\propto a^{-3.5}$ within the
  size interval of $[0.001,0.5]$~$\mu$m. It is
  hereafter referred as the `MRN' distribution.
\item We investigate three values of $f_d$: $0.001$, $0.01$ and
  $0.1$. The two extreme values roughly bracket the current
  theoretical and observational constraints on dust abundance in the
  ICM/IGM (i.e $10.^{-5}$ and $10^{-3}$ in terms of dust-to-gas mass
  ratio) \citep{popescu00,aguirre01,chelouche07,muller08,giard08}.
\end{itemize}

Tab.~\ref{tab:simparam} lists code names and simulation details of all
runs used in this work. In case of models D1 to D5, simulation cooling
rates are given by the added effect of cooling due to dust and
radiative gas cooling.
Total cooling functions are displayed (non-coloured lines) in the
bottom panel of Fig~\ref{fig:lamdust} for each of these models at
$Z/Z_\odot=0.33$.  As the Figure indicates, the effect of dust cooling
is stronger for models with higher dust-to-metal mass abundance
parameters, $f_d$, and for smaller grain sizes (model D1). For low
values of $f_d$ the impact of dust cooling is significantly
reduced. For example, in the case of model D5, the contribution of
dust to the total cooling rate is negligible at $Z/Z_\odot=0.33$ for
all temperatures. Therefore we do not expect to obtain significant
differences between simulations with these two models.

\section{Numerical Simulations}
\label{sec:sinum}

\subsection{Simulation description}

Simulations were carried out with the public code package {\tt Hydra},
\citep{couchman:1995, pearce:1997}, an adaptive
particle-particle/particle-mesh (AP$^3$M), \citep{couchman:1991}
gravity solver with a formulation of smoothed particle hydrodynamics
(SPH), see \citet{thacker2000}, that conserves both entropy and
energy. In simulations with cooling gas particles are allowed to cool
using the method described in \citet{thomas:1992} and the cooling
rates presented in previous Section.  At a given time step, gas
particles with overdensities (relative to the critical density) larger
than $10^4$, and temperatures below $1.2\times 10^4$K are converted
into collisionless baryonic matter and no longer participate in the
gas dynamical processes. The gas metallicity is assumed to be a global
quantity that evolves with time as $Z=0.3(t/t_0)Z_\odot$, where
$Z_\odot$ is the solar metallicity and $t/t_0$ is the age of the
universe in units of the current time.

All simulations were generated from the same initial conditions
snapshot, at $z=49$.  The initial density field was constructed, using
$N=4,096,000$ particles of baryonic and dark matter, perturbed from a
regular grid of fixed comoving size $L=100 \, h^{-1} {\rm Mpc}$.  We
assumed a $\Lambda$-CDM cosmology with parameters, $\Omega =0.3$,
$\Omega_{\Lambda}=0.7$, $\Omega_{b}=0.0486$, $\sigma_{8}=0.9$,
$h=0.7$. The amplitude of the matter power spectrum was normalized
using $\sigma_8=0.9$. The matter power spectrum transfer function was
computed using the BBKS formula \citep{bardeen:1986}, with a shape
parameter $\Gamma $ given by the formula in
\citet{sugiyama:1995}. With this choice of parameters, the dark matter
and baryon particle masses are $2.1\times 10^{10} \, h^{-1} {\rm
M_{\odot}}$ and $2.6 \times 10^{9} \, h^{-1} {\rm M_{\odot}}$
respectively. The gravitational softening in physical coordinates was
$25\,h^{-1} {\rm kpc}$ below $z=1$ and above this redshift scaled as
$50(1+z)^{-1}\,h^{-1} {\rm kpc}$.

\begin{table}
\begin{center}
\vspace{0.3cm}
\begin{tabular}{lcccc}
\hline \hline
Run   & Physics             & $f_d$ & Grain size & $N_{\rm steps}$\\ 
\hline \hline
A & adiabatic (no dust) & -            & -      &             2569   \\
C & cooling (no dust)   & -            & -      &             2633   \\
\hline
D1 & cooling with dust   & 0.100        & small  &             2944   \\
D2 & cooling with dust   & 0.100        & MRN    &             2920   \\
D3 & cooling with dust   & 0.100        & big    &             2886   \\
D4 & cooling with dust   & 0.010        & MRN    &             2698   \\
D5 & cooling with dust   & 0.001        & MRN    &             2633   \\
\hline \hline
\end{tabular}
\caption[Simulation runs]{Simulation parameters: $f_d$, dust-to-metal
mass ratios (see Eq.~\ref{eq:zd}), grain sizes, and number of
timesteps taken by simulation runs to evolve from z=49 to
z=0. Cosmological and simulation parameters were set the same in all
simulation, as follows: $\Omega =0.3$, $\Omega_{\Lambda}=0.7$,
$\Omega_{b}=0.0486$, $\sigma_{8}=0.9$, $h=0.7$, boxsize $L=100 \,
h^{-1} {\rm Mpc}$, and number of baryonic and dark mater particles,
$N=4,096,000$.
\label{tab:simparam}
}
\end{center}
\end{table}

We generate a total of 7 simulation runs, listed in
Table~\ref{tab:simparam}. The first two runs, which will be referred
hereafter as `adiabatic' (or model 'A') and `cooling' (or model 'C')
simulations, do not include dust. Simulations 3 to 7 differ only on
the dust model parameters assumed in each case, and will be referred to
as `dust' runs, and are labeled as 'D1' to 'D5' models (see
Sec.~\ref{sec:impmod} for details on the dust models definition). 
This will allow us to investigate the effects of the dust model
parameters on our results. The last column in the table gives the total
number of timesteps required by each simulation to arrive to redshift
zero. For each run we stored a total of 78 snapshots in the redshift
range $0<z<23.4$. Individual snapshots were dump at redshift intervals
that correspond to the light travel time through the simulation box,
ie simulation outputs stack in redshift.

\subsection{Catalogue construction}

Cluster catalogues are generated from simulations
using a modified version of the Sussex extraction software
developed by Thomas and collaborators \citep{thomas:1998,
pearce:2000,muanwong:2001}. Briefly, the cluster identification
process starts with the creation of a minimal-spanning tree of dark
matter particles which is then split into clumps using a maximum
linking length equal to $0.5\,\Delta_{\rm b}^{-1/3}$ times the mean
inter-particle separation. Here $\Delta_{\rm b}$ the contrast
predicted by the spherical collapse model of a virialized sphere
\citep{eke:1998}. A sphere is then grown around the densest dark
matter particle in each clump until the enclosed mass verifies
\begin{equation}
M_\Delta(<R_\Delta)=\frac{4\pi}{3}R^3_\Delta\,\Delta\,\rho_{\rm crit}(z).
\label{eq:mass}
\end{equation}
where $\Delta$ is a fixed overdensity contrast, $\rho_{\rm
crit}(z)=(3H_0^2/8\pi G)E^2(z)$ is the critical density and
$E(z)=H(z)/H_0=\sqrt{(\Omega(1+z)^3+\Omega_\Lambda}$. Cluster
properties are then computed in a sphere of radius $R_{200}$, ie with
$\Delta=200$, for all objects found with more than 500 particles of
gas and dark matter. This means that our original catalogues are
complete in mass down to $1.18\times10^{13}h^{-1}M_{\sun}$.  For the
study presented in this paper we have trimmed our original catalogues
to exclude galaxy groups with masses below $M_{\rm lim}=
5\times10^{13}h^{-1}M_{\sun}$. In this way the less massive object
considered in the analysis is resolved with a minimum of 2100
particles of both gas and dark mater. Our catalogues at z=0 have at
least 60 clusters with masses above $M_{\rm lim}$. This number drops
to about 20 clusters at z=1.

Cluster properties investigated in this paper are the mass, $M$,
mass-weighted temperature, $T_{\rm mw}$ and entropy, $S$ 
(defined as $S=k_{\rm B}T/n^{-2/3}$), 
integrated Compton parameter, $Y$ (i.e roughly the
SZ signal times the square of the angular diameter distance to the
cluster), and core excised (50 $h^{-1}$kpc) X-ray bolometric
luminosity, $L_{\rm X}$. These were computed in the catalogues
according to their usual definitions, see \citet{dasilva2004}:

\begin{equation}
M=\sum_k m_k,
\label{eqn:m}
\end{equation}
\begin{equation}
T_{\rm mw} = { \sum_{i} m_i \, T_i \over \sum_{i} m_i},
\label{eqn:tmw}
\end{equation}
\begin{equation}
S = { \sum_{i} m_i \, k_{\rm B}T_i\, n_i^{2/3} \over \sum_{i} m_i},
\label{eqn:s}
\end{equation}
\begin{equation}
Y = \frac{k_{\rm B}\sigma_{\rm T}}{m_{\rm e}c^2}\frac{(1+X)}{2m_{\rm H}}
\, \sum_{i}{m_i \, T_i}, 
\label{eq:y}
\end{equation}
\begin{equation}
L_{\rm X} =  \sum_{i} 
              {m_i \, \rho_i \, \Lambda_{\rm bol}(T_i,Z)
              \over (\mu m_{\rm H})^2 },
\end{equation}
where summations with the index {\it i} are over hot ($T_i > 10^5$K)
gas particles and the summation with the index {\it k} is over all
(baryon and dark matter) particles within $R_{200}$. Hot gas is
assumed fully ionised. The quantities $m_i$, $T_i$, $n_i$ and $\rho_i$
are the mass, temperature, number density and mass density of gas
particles, respectively. $\Lambda_{\rm bol}$ is the bolometric cooling
function in \citet{sutherland:1993} and $Z$ is the gas metallicity.
Other quantities are the Boltzmann constant, $k_{\rm B}$, the Thomson
cross-section, $\sigma_{{\rm T}}$, the electron mass at rest, $m_{{\rm
e}}$, the speed of light $c$, the Hydrogen mass fraction, $X=0.76$,
the gas mean molecular weight, $\mu $, and the Hydrogen atom mass,
$m_{\rm H}$.

\section{Scaling Relations}
\label{sec:scal}
In this paper we investigate the scalings of mass-weighted
temperature, $T_{\rm mw}$, entropy, $S$, integrated Compton parameter,
$Y$ and core excised X-ray bolometric luminosity, $L_{\rm X}$, with
mass, $M$. Taking into account Eq.~(\ref{eq:mass}) these cluster
scaling relations can be expressed as:
\begin{equation}
T_{\rm mw}=A_{\rm TM}\,(M/M_0)^{\alpha_{\rm TM}}\,(1+z)^{\beta_{\rm TM}}\, E(z)^{2/3}  \,,
\label{eq:tm}
\end{equation}
\begin{equation}
S=A_{\rm SM}\,(M/M_0)^{\alpha_{\rm YT}}\,(1+z)^{\beta_{\rm YT}}\, E(z)^{-2/3} \,,
\label{eq:sm}
\end{equation}
\begin{equation}
Y=A_{\rm YT}\,(M/M_0)^{\alpha_{\rm YM}}\,(1+z)^{\beta_{\rm YM}}\, E(z)^{2/5} \,,
\label{eq:ym}
\end{equation}
\begin{equation}
L_{\rm X}=A_{\rm LM}\,(M/M_0)^{\alpha_{\rm LM}}\,(1+z)^{\beta_{\rm LM}}\, E(z)^{7/3} \,,
\label{eq:lxm}
\end{equation}
where $M_0=10^{14}h^{-1}{\rm M_{\odot}}$ and the powers of the $E(z)$
give the predicted evolution, extrapolated from the self-similar
model, \citep{kaiser1986}, of the scalings in each case. The
quantities, $A$, $\alpha$, and $\beta$, are the scalings normalisation at
$z=0$; the power on the independent variable; and the departures from
the expected self similar evolution with redshift. 

These scalings can be expressed in a condensate form,
\begin{equation}
y\, f(z)=y_0(z) \,(x/x_0)^\alpha  \,, 
\label{eqyx}
\end{equation}
where $y$ and $x$ are cluster properties (e.g. $T_{\rm mw}$, $M$), 
\begin{equation}
y_0(z) = A \, (1+z)^\beta \,,
\label{eqy0z}
\end{equation}
and $f(z)$ is some fixed power of the cosmological factor $E(z)$.
To determine $A$, $\alpha$, and $\beta$ for each scaling we use the
method described in \citet{dasilva2004, aghanim2008}. To summarize, the
method involves fitting the simulated cluster populations at each
redshift with Eqs.~(\ref{eqyx}) and (\ref{eqy0z}) written in
logarithmic form. First we fit the cluster distributions with a
straight-line in logarithmic scale at all redshifts.
If the logarithmic slope $\alpha$ remains approximately
constant (i.e. shows no systematic variations) within the redshift
range of interest, we then set $\alpha$ as the best fit
value at $z=0$. 
Next, we repeat the fitting procedure with $\alpha$ fixed to
$\alpha(z=0)$ to determine the scaling normalisation factors
$y_0(z)$. This avoids unwanted correlations between $\alpha$ and
$y_0(z)$. The r.m.s. dispersion of the fit is also
computed at each redshift according to the formula,
\begin{equation}
\sigma_{\log y'} =\sqrt{ {1 \over N} \sum_{i} (\log (y'_i/y'))^2} \,,
\label{sigmay}
\end{equation}
where $y'=yf$ (see Eq.~(\ref{eqyx})) and $y'_i$ are individual data
points. 
Finally, we perform a linear fit of the
normalisation factors with redshift in logarithmic scale, see
Eq.~(\ref{eqy0z}), to determine the parameters $A$ and $\beta$.

We note that above $z=1.5$ the number of clusters in our catalogues
decreases typically below 10, hence, we do not fit the scaling
relations above this redshift value. 

\begin{figure*}
\begin{center}  
\epsfxsize=8.30cm \epsfbox{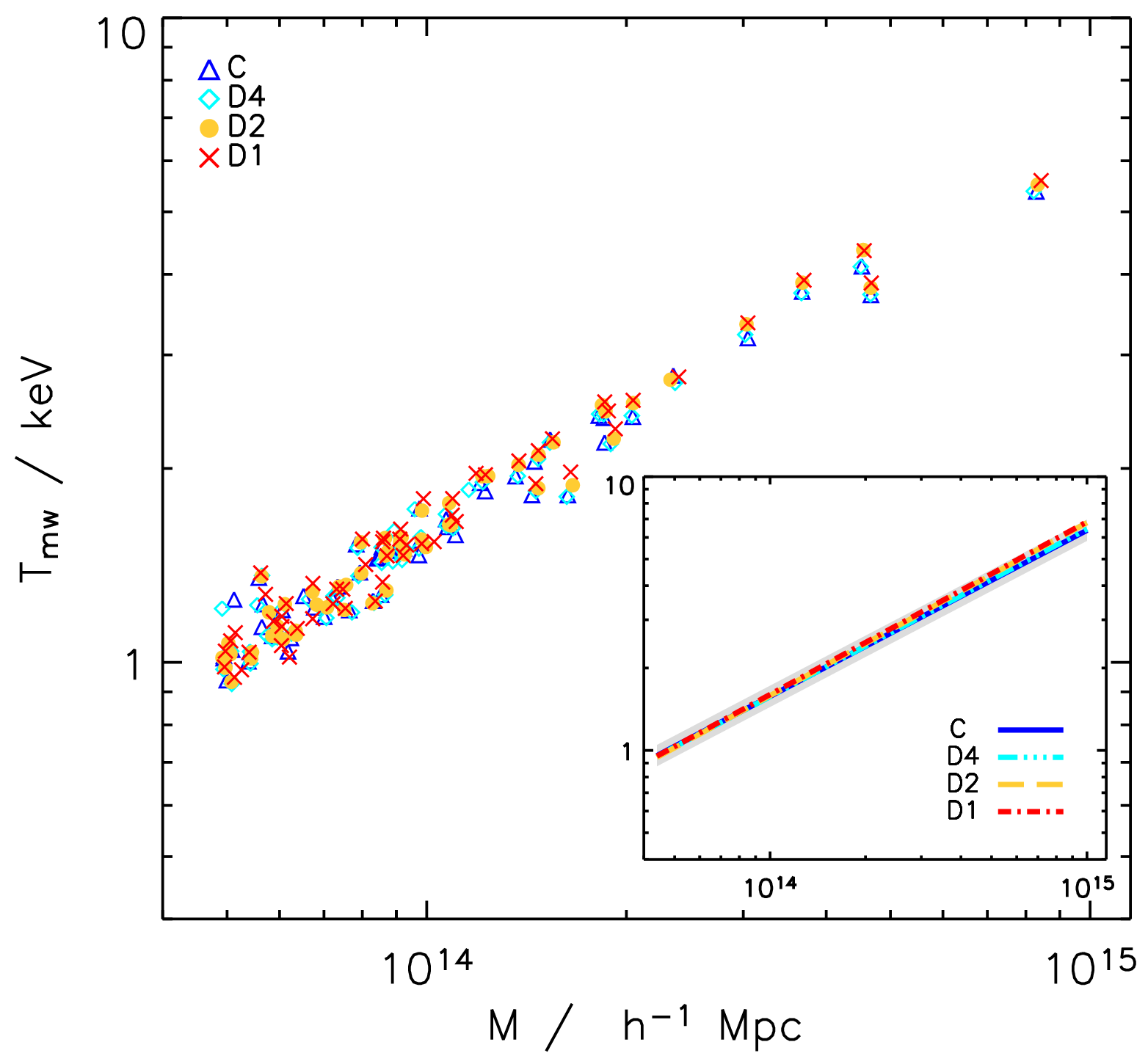} \hspace{0.9cm} 
\epsfxsize=8.30cm \epsfbox{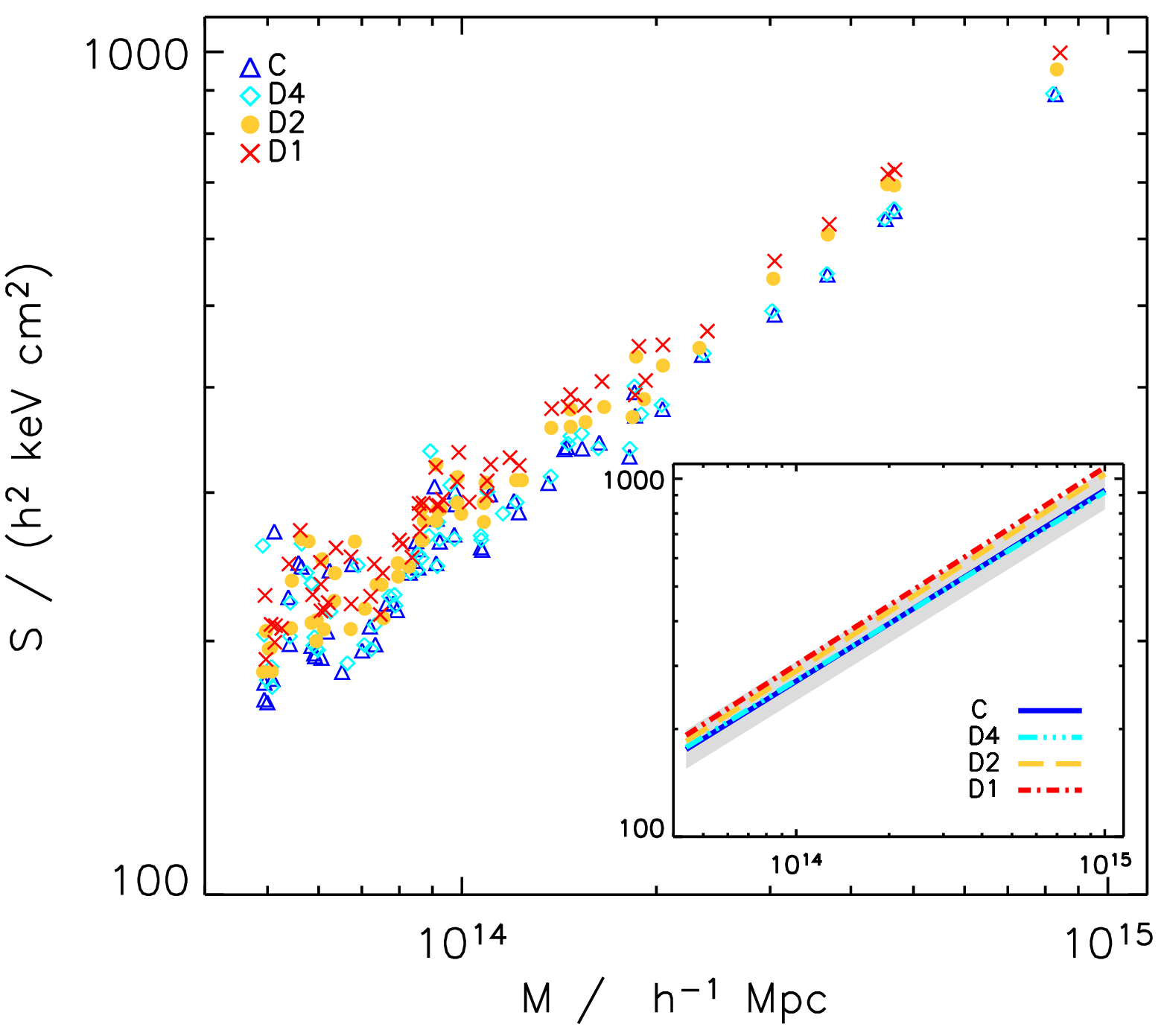}\\
\epsfxsize=8.30cm \epsfbox{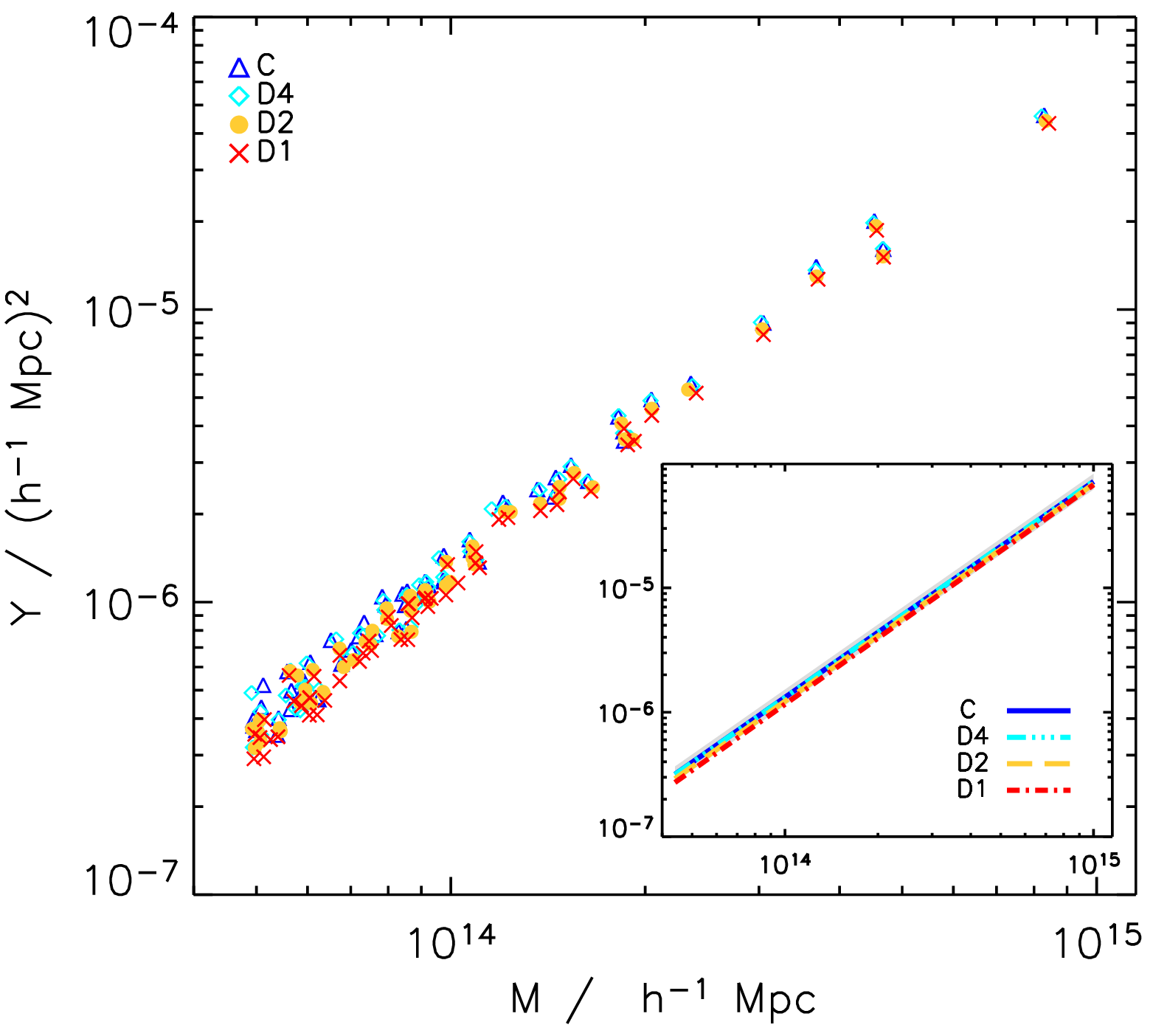} \hspace{0.9cm}
\epsfxsize=8.30cm \epsfbox{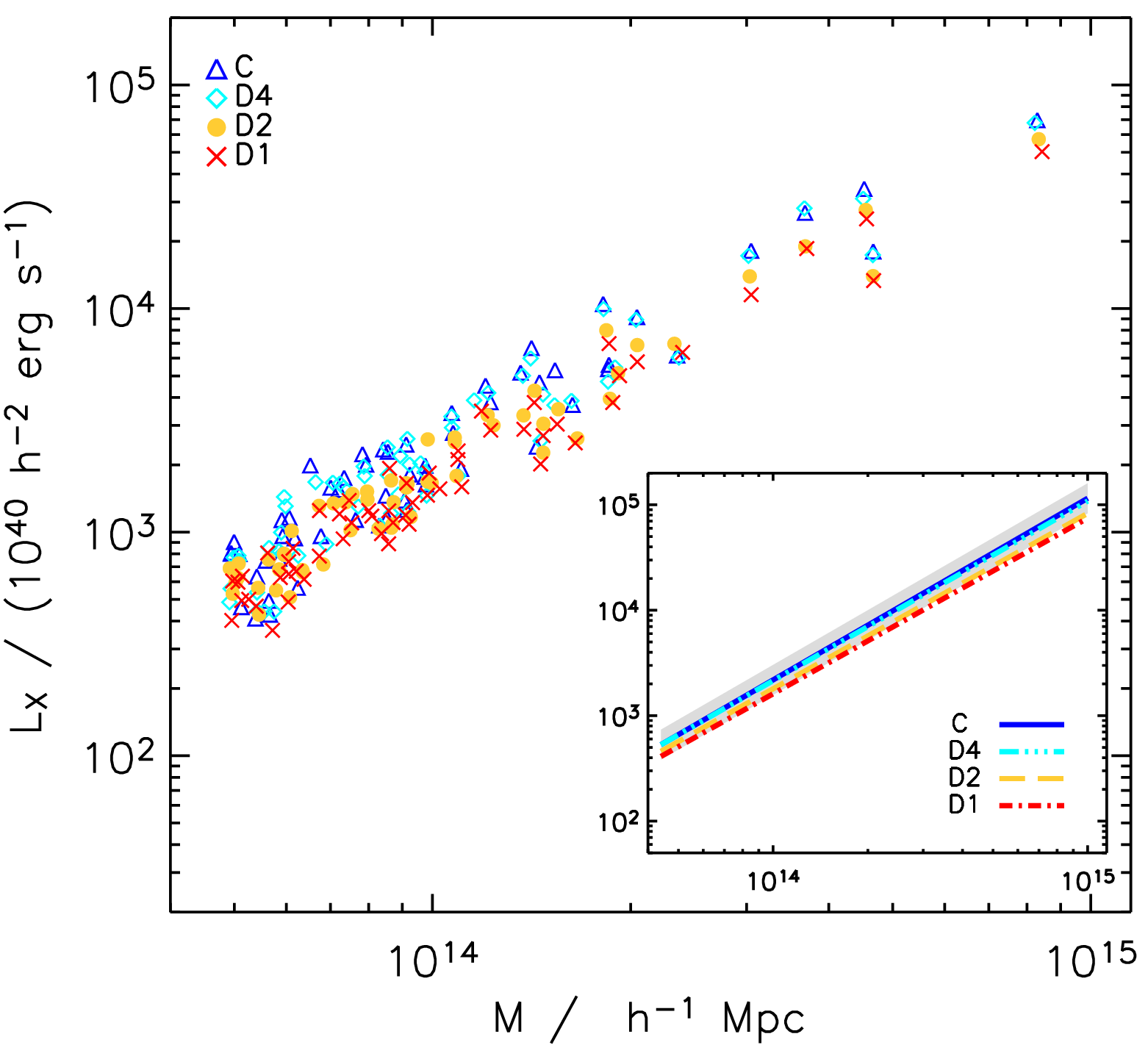}
\caption{\label{fig:scalings_z=0} Cluster scalings at redshift zero
for the $T_{\rm mw}-M$ (top left panel), $S-M$, (top right panel),
$Y-M$ (bottom left panel), and $L_{\rm X}-M$ (bottom right
panel). Displayed quantities are computed within $R_{\rm 200}$, the
radius where the mean cluster density is 200 times larger than the
critical density.  Blue colour and triangles stand for the cooling
(C) run, cyan and diamonds are for the D4 run, yellow and filled
circles are for clusters in the D2 run, and red and crosses are for
the D1 run. The lines in the embedded plots are the best-fit lines to
the cluster distributions and the shaded areas are the
fit r.m.s. dispersions for the C model, for each scaling.  
}
\end{center}
\end{figure*}

\section{Results}
\label{sec:res}
 
\subsection{Scaling relations at $z=0$}

In this section we present cluster scaling relations obtained from
simulations at redshift zero. We investigate the four scalings
presented in Section~\ref{sec:scal} for all models under
investigation.  

Figure~\ref{fig:scalings_z=0} shows the $T_{\rm mw}-M$ (top left
panel), $S_{\rm mw}-M$, (top right panel), $Y-M$ (bottom left panel),
and $L_{\rm X}-M$ (bottom right panel) scalings, with all quantities
computed within $R_{\rm 200}$. In each case, the main plot shows the
cluster distributions for the C (triangles), D4 (diamonds), D2 (filled
circles) and D1 (crosses) simulations, whereas the embedded plot
presents the power law best fit lines (solid, triple dot-dashed,
dashed and dot-dashed for C, D4, D2 and D1 models, respectively)
obtained in each case, colour coded in the same way as the cluster
distributions. Here we have chosen to display dust models that allow
us to assess the effect of dust parameters individually. For example,
the dust models in runs D4 and D2 only differ by the dust-to-metal
mass ratio parameter, whereas models D2 and D1 have different grain
sizes but the same $f_d$. The shaded gray areas in the embedded plots
give the r.m.s. dispersion of the fit for cooling (C) model.
The dispersions obtained for the other models have similar
amplitudes to the C case. The scalings of entropy and X-ray luminosity
with mass show larger dispersions because they are more sensitive to
the gas physical properties (density and temperature) in the inner
parts of clusters than the mass-weighted temperature and $Y$
versus mass relations which are tightly correlated with mass. 

An inspection of Fig.~\ref{fig:scalings_z=0} allows us to conclude
that the cluster scalings laws studied here are sensitive to the
underlaying dust model, and in particular to models where the dust
cooling is stronger (model D1 and D2).  The differences are more
evident in the $S-M$ and $L_X-M$ scalings, but are also visible, to a
lower extent, in the $T_{\rm mw}-M$ and $Y-M$ relations. 
Generally, the inclusion of dust tends to increase temperature and
entropy because the additional cooling increases the formation of
collisionless (star forming) material leaving the remaining particles
in the gas phase with higher mean temperatures and entropies. The
decrease of $Y$ and X-ray luminosities reflects the
effect of lowering the hot gas fraction and density due to dust
cooling. These effects dominate over the effect of increasing the
temperature.

In fact a closer inspection of Fig.~\ref{fig:scalings_z=0} indicates
that differences for the same cluster in different models (note that
all simulations have the same initial conditions so a
cluster-to-cluster comparison can be made), reflect the differences of
intensity between cooling functions presented in
Figure~\ref{fig:lamdust}. For example, the differences between models
D4 and C are clearly small as one would expect from the small
differences between cooling functions displayed in the bottom panel of
Figure~\ref{fig:lamdust}. Another interesting example is that an
increase of one order of magnitude in $f_d$ from D4 to D2 seems to
cause a stronger impact in the properties of the most massive clusters
than the differences arising from changing the dust grain sizes from
D2 to D1. Again this reflects the differences between cooling functions,
which in the latter case are smaller at higher temperatures (see
bottom panel of Figure~\ref{fig:lamdust}).

A way of quantifying the effect of dust, is to look at the best fit
slope, $\alpha$, and normalisation, $\log A$, parameters of these
scalings which are presented in Table~\ref{tab_best_fit} for all
cooling models considered in this paper.
We find that fitting parameters are quite similar for models C, D5,
and D4 whereas models with high dust abundances provide the strongest
variations of the fitting parameters, particularly for the
normalisations. In several cases, differences are larger than the
(statistical) best-fit errors, particularly for the D1 and D2 models.
We also investigated scalings at redshift zero for the A (adiabatic)
model and found they were consistent with self-similar predictions.
As expected, the results obtained for the adiabatic and cooling models
are in very good agreement with the findings of \citep{dasilva2004,
aghanim2008} which use similar simulation parameters and cosmology.

\subsection{Evolution of the scaling relations}
 
We now turn to the discussion of the evolution of the cluster scaling
laws in our simulations. Here we apply the fit to a power law
procedure described in Section~\ref{sec:scal} to derive the
logarithmic slope, $\beta $, of our fitting functions,
Eqs.~(\ref{eq:tm})-(\ref{eq:lxm}). As mentioned earlier, this quantity
measures evolution departures relative to the self-similar
expectations for each scaling.

%
%
\begin{figure}
\begin{center} 
\epsfxsize=8.5cm \epsfbox{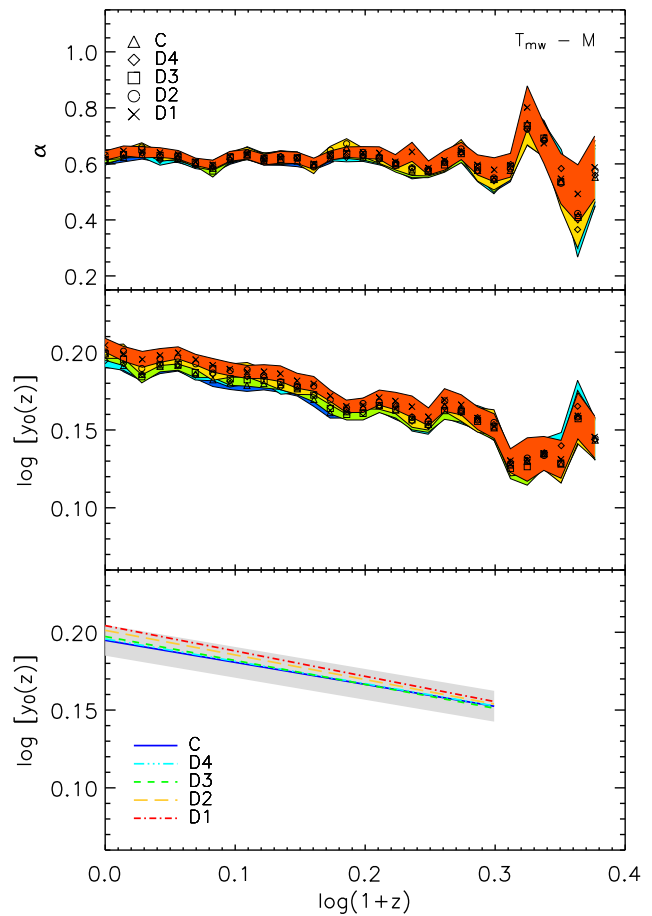}
\caption{\label{fig:tmw-M_zevol} Evolution of the slope (top panel),
normalisation (middle panel), and normalisation best
fit lines (bottom panel) of the $T_{\rm mw}-M$ cluster scaling
relation for the C (triangles, solid line), D4 (diamonds,
triple-dot-dashed line), D3 (squares, short-dashed line), D2 (circles,
dashed line) and D1 (crosses, dot-dashed line) simulation
models. Colour bands are best fit errors to the cluster distributions
at each redshift. The shaded area in the bottom panel is the rms
dispersion of the normalisation fit for the cooling model.(see text
for details) }
\end{center}
\end{figure}
%

%
%
\begin{figure}
\begin{center}
\epsfxsize=8.5cm \epsfbox{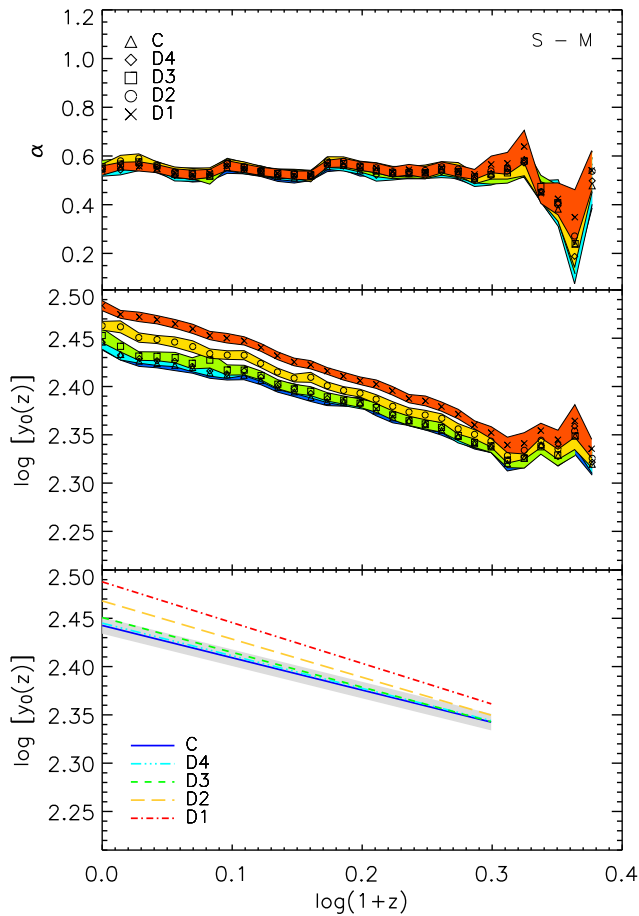}
\caption{\label{fig:S-M_zevol} Evolution of the slope (top panel),
normalisation (middle panel), and normalisation best fit lines (bottom
panel) of the $S-M$ cluster scaling relation for the C, D4, D3, D2, D1
simulation models. Symbols, lines and colours are the same as in
Fig~\ref{fig:tmw-M_zevol}.
}
\end{center}
\end{figure}

\begin{figure}
\begin{center}
\epsfxsize=8.5cm \epsfbox{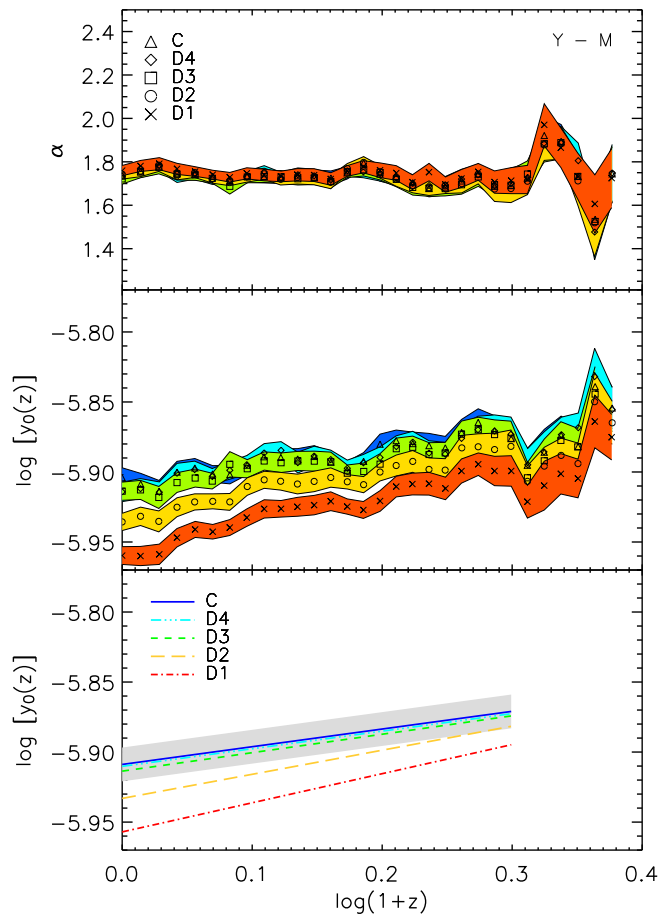}
\caption{\label{fig:Y-M_zevol} 
Evolution of the slope (top panel),
normalisation (middle panel), and normalisation best fit lines (bottom
panel) of the $Y-M$ cluster scaling relation for the C, D4, D3, D2, D1
simulation models. Symbols, lines and colours are the same as in
Fig~\ref{fig:tmw-M_zevol}.
}
\end{center}
\end{figure}

\begin{figure}
\begin{center}
\epsfxsize=8.5cm \epsfbox{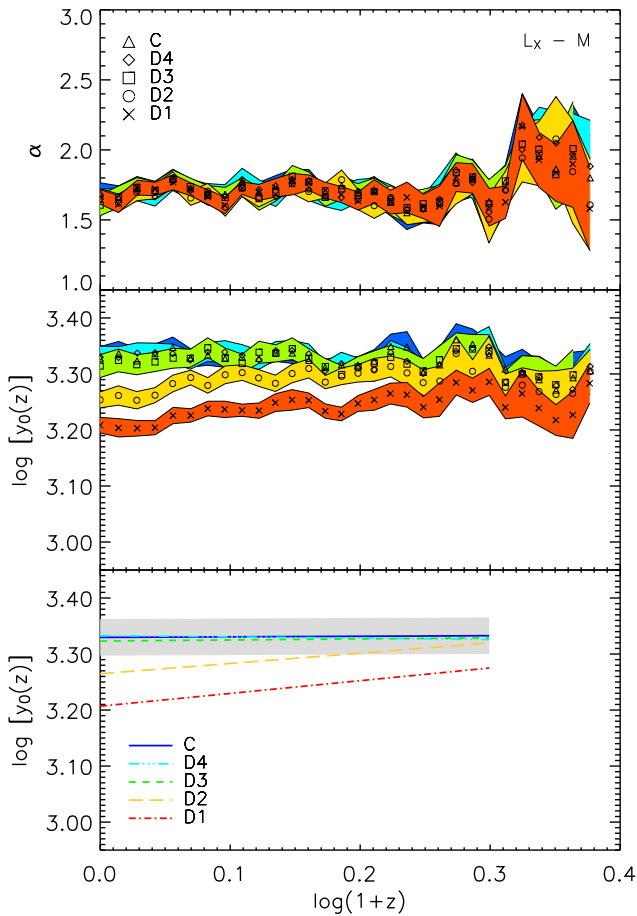}
\caption{\label{fig:Lxb-M_zevol} 
Evolution of the slope (top panel),
normalisation (middle panel), and normalisation best fit lines (bottom
panel) of the $L_{X}-M$ cluster scaling relation for the C, D4, D3, D2, D1
simulation models. Symbols, lines and colours are the same as in
Fig~\ref{fig:tmw-M_zevol}.
}
\end{center}
\end{figure}

In Figs.~\ref{fig:tmw-M_zevol}, \ref{fig:S-M_zevol},
\ref{fig:Y-M_zevol}, and \ref{fig:Lxb-M_zevol} we plot the redshift
dependence of the power law slopes, $\alpha$, (top panels), and
normalisations, $\log y_0$, (middle panels) for our $T_{\rm mw}-M$,
$S-M$, $Y-M$, and $L_{\rm X}-M$ scalings, respectively. The bottom
panels show straight lines best fits, up to z=1, to the data points in
the middle panels of each Figure. The slopes of these lines are the
$\beta$ parameters in Eqs.~(\ref{eq:tm})-(\ref{eq:lxm}). We decided
not to include data points above z=1 in the computation of $\beta $
because cluster numbers drop rapidly (below 20) which, in some cases,
causes large oscillations in the computed normalizations. Moreover
in the case of the $L_X-M$ relation, the evolution of $y_0(z)$ with
redshift appears to deviate from a straight line above $z\simeq 1$.
In Table~\ref{tab_best_fit} we provide a complete list of the $\log
A$, $\beta $ and $\alpha$ fitting parameters and associated
statistical errors for all scalings and cooling models investigated in
this paper. The displayed values are valid in the redshift range
$0<z<1$. In the top and middle panels the coloured bands correspond to
the $\pm 1\sigma$ envelope of the best fit errors obtained at each
redshift for $\alpha$, and $\log y_0$. The shaded area in the bottom
panels are r.m.s. fit dispersions of the normalisations, $\log y_0$,
computed for the cooling model using Eq.~(\ref{sigmay}).

Results from different simulation runs are coded in the following way:
triangles and solid lines stand for the cooling model, diamonds and
triple-dot-dashed lines represent the D4 model, squares and
short-dashed lines are for D3 model, circles and dashed lines for the
D2 models and crosses and dot-dashed lines are for the D1 model. Here
we have omitted the D5 model for clarity. It provides the same fitting
results as the cooling model. This confirms expectations and the
comments made in the last paragraph of Section~\ref{sec:impmod}.

The top panels of these Figures allow us to conclude that the $\alpha$
slopes of our scalings are fairly insensitive to dust
cooling. These also show no evidence of systematic variations with
redshift for all scalings, which is an important requirement when
fitting the cluster distributions with power-laws of the form
Eqs.~(\ref{eq:tm})-(\ref{eq:lxm}). The redshift independence of the
slopes with the dust model confirms our findings at redshift
zero. The scatter (non-systematic ``oscillations'') at high redshift is
caused by the decrease of the number of clusters with $M_{\rm lim}\ge
5\times10^{13}h^{-1}M_{\sun}$, the sample selection used for all fits.

The main effect of cooling by dust is reflected in the changes it
produces in the normalisations of the cluster scaling laws. Again, the
the impact of dust is different depending on the scaling under
consideration. For the $T_{\rm mw}-M$ scaling in
Fig.~\ref{fig:tmw-M_zevol} we see a sytematic variation with the dust
model (ordered in the following way: C, D4, D3, D2, D1), but
differences between models are within the errors and fit dispersions
of each others. For the evolution of the normalisations of the $S-M$,
$Y-M$, and $L_{\rm X}-M$ scalings (see Figs.~\ref{fig:S-M_zevol},
\ref{fig:Y-M_zevol}, and \ref{fig:Lxb-M_zevol}) we conclude that the
inclusion of dust cooling causes significant departures from the
standard radiative cooling model depending on the dust model
parameters. For example, this is clear from the non-overlapping errors
and fit dispersions of the normalisations for the D2 and D1
models. For all scalings, the relative strength of the effect of dust
follows the relative intensity of the cooling functions presented in
Section~\ref{sec:impmod}. This orders the models in the following
way: C, D4, D3, D2, D1 with increasing normalisations for the $T_{\rm
mw}-M$ and $S-M$ scalings and decreasing normalisations for the $Y-M$
and $L_X-M$ relations. 
 
We end this section by noting that we find positive evolution
(relative to the expected self-similar evolution, i.e. for a given $x$
in Eq~(\ref{eqyx}) the property $yf$ is higher at higher redshifts)
for the $Y-M$ and $L_X-M$ (models D1 and D2 only) relations whereas the
$T_{\rm mw}-M$, and $S-M$ relations show negative evolutions
relative to the self-similar model. This is in line with the findings
from simulations with radiative cooling of similar size and cosmology,
see eg \citep{dasilva2004, aghanim2008}.

%
%
%

%
%

\section{discussion}
\label{sec:dis}

\subsection{Efficiency of the dust cooling}
 
In agreement with the cooling functions of \citep{montier04}, the dust
cooling is most effective in the cluster temperature regime. The
relative importance of the dust cooling with respect to the gas
radiative cooling is strongly dependent on the dust abundances and the
intrinsic physical properties of the dust.  This is clearly shown in
our scaling relations results:

\begin{itemize}
\item the $T_{\rm mw}-M$ relation is almost unchanged when adding dust
  cooling to the radiative gas cooling (see
  Fig.~\ref{fig:tmw-M_zevol}). Our results show that the (mass
  weighted) temperature--mass relation within $R_{200}$, is
  essentially driven by the gravitational heating of the gas (due to
  its infall on the cluster potential well), and that the physics of
  baryons (at least for the physics implemented in the simulations
  presented in this paper) play very little role in the outer parts of
  halos which dominate the estimation of the mass-weighted temperature and
  integrated mass.
  Since gas cooling tends to disturb the dark matter distribution at
  the centre of clusters in high resolution simulations
  \citep{gnedin04}, the cooling by dust may amplify this effect, and
  thus modify scaling relations like the $T_{\rm mw}-M$. In the case
  of observationally derived quantities, scaling laws will be drawn
  from overall quantities that will proceed from mixed-projected
  information over a wide range of radii. If a gradient exist in the
  dust effect towards the cluster centre, an ``overall'' temperature
  might bear the signature of the structural effect of dust. Anyway
  this quantification is beyond the scope of this paper and will be
  investigated in a forthcoming paper. There is also no significant
  effect between the different dust models and the radiative case on
  the evolution of the slope and normalisation of the $T_{\rm mw}-M$
  relation.

\item On the other hand the other three scaling laws are deeply
  related to the clusters baryonic component. The clear effect on the
  $S-M$, $Y-M$, and $L_{\rm X}-M$ relations illustrates this fact (see
  Figs.~\ref{fig:S-M_zevol}, \ref{fig:Y-M_zevol}, and
  \ref{fig:Lxb-M_zevol}). We found that the slopes of these scalings
  remain fairly insensitive to dust, whereas normalisations show
  significant changes depending on the dust parameters. Relative
  changes in the normalisations at redshift zero and
  $M_0=10^{14}h^{-1}{\rm M_{\odot}}$ can be as high as 25\% for
  $L_{\rm X}-M$ and 10\% for the $S-M$, $Y-M$ relations for the D1
  model. Models with lower dust abundances and MRN grain size
  distributions present smaller but systematic variations relative to
  the C model. As any other cooling process, the cooling due to
  dust tends to lower the normalisations of the $Y-M$, and $L_{\rm
  X}-M$ scalings due to the decrease of the hot gas fractions and
  densities which dominate the increase of temperature. The increase
  of normalisations for the $S-M$ and $T_{\rm mw}-M$ relations with
  added dust cooling is also in line with expectations because cooling
  converts cold, dense, gas into collisionless star forming material,
  which raises the mean temperature and entropy of the remaining gas.

  \item Our simulations allow us to quantify the relative impact of
  the dust parameters on the investigated cluster scalings (see
  Figs.~\ref{fig:tmw-M_zevol} to \ref{fig:Lxb-M_zevol} and
  Table~\ref{tab_best_fit}). From one model to another one can
  identify two clear effects due to dust: (i) it shows the expected
  effect of the dust abundance, which from models D4 to D2 raise by a
  factor of 10, producing a change of normalisations relative to the
  purely radiative case (model C), from almost zero percent
  contribution to about 14\%, 5\% and 6\% contributions for the
  $L_{\rm X}-M$, $Y-M$ and $S-M$ relations, respectively. (ii) Even
  more striking is the effect of the intrinsic dust grain physical
  properties. The variation of normalisations relative to the C model,
  change from a zero percent level for model D4 to about 25\% ($L_{\rm
  X}-M$) and 10\% ($Y-M$, and $S-M$) for the model D1 (ie the relative
  change from models D2 to D1 is about 13\% and 5\%,
  respectively). All these percentages were calculated using
  normalisations at redshift zero and $M_0=10^{14}h^{-1}{\rm
  M_{\odot}}$. Therefore the size of the grains comes to be an equally
  important parameter varying the efficiency of the dust
  cooling. The smaller the grain, the stronger the cooling.

  \item From Figs.~\ref{fig:tmw-M_zevol}, \ref{fig:S-M_zevol},
  \ref{fig:Y-M_zevol}, and \ref{fig:Lxb-M_zevol} one finds that
  differences between normalisations become progressively important
  with decreasing redshift. This confirms expectations because
  metallicity was modelled in simulations as a linearly increasing
  function of time. Although our implementation of metallicity should
  only be regarded as a first order approximation to the modelling of
  more complex physical processes (acting on scales below the
  resolution scale of the present set simulations), it would be
  interesting to investigate whether a similar effect remains (ie the
  effects of dust become progressively important at low redshift) when
  such processes are taken into account throughout the formation
  history of galaxy clusters (see discussion below).

\end{itemize}~\\[-2em]

\subsection{Limitation of the dust implementation}
\label{sec:limit}
In order to implement the presence of dust in the numerical
simulations, we chose a ``zero order approach'': we directly
correlated the presence of dust with the presence of metals under the
assumption that there is no segregation in the nature of the material
withdrawn from galaxies and injected in the IGM/ICM (metals, gas,
stars or dust). However, this assumption suffers from limitations
linked to the dust lifetime. Indeed, dust grains strongly suffer of
sputtering and within their lifetime they are depleted from metals
which, contrary to dust grains, are not destroyed i.e.  remain in the
ICM/IGM. Therefore our whole analysis is to be considered within the
framework of this assumption, and is to be understood as a basic
implementation of the effects of dust with the objective of assessing
whether dust has a significant impact on large scale structure
formation, and consequently to quantify these effects at first order.

Moreover, our implementation is also {\it ad hoc}. Indeed, beside the
cooling function of dust, our implementation is not a physical
implementation. We did not deal {\it stricto senso} with the physics
of the dust creation and dust destruction processes. This would be a
step further, and is yet beyond the scope of this paper as mentioned
above. However, making use of the cooling function by
\citep{montier04}, we have performed a fully self-consistent
implementation of the effect of dust as a cooling vector of the
ICM/IGM. Indeed, on the basis of the cooling function, the
implementation encapsulates the major physical processes to which dust
is subjected and acts as a non-gravitational process at the scale of
the ICM and the IGM.

As already mentioned, we directly correlated the abundance of dust
with metallicity, thus to the metallicity evolution, which chosen
evolution law is quite drastic: $Z=0.3(t/t_0)Z_\odot$. Indeed, if the
metallicity at $z=0$ is normalized to the value of $0.3Z_\odot$, it is
lowered to $\sim 0.2$ at $z=0.5$ and $\sim 0.1$ at $z=1$. However,
other numerical works based on simulations including physical
implementation of metal enrichment processes but without dust agree
well with observational constraints (mainly provided by X-ray
observations of the Fe-K line) which indicate high metalicity values,
$Z\sim 0.3 Z_{\odot}$, up to redshifts above 1.0
\citep{cora08,borgani08}. This shows that, as for the stellar
component which is already in place in galaxies when clusters form,
the metal enrichment of the ICM/IGM has occured through the feedback
of galaxy formation and evolution, and therefore it {\it de facto}
strongly enriched the IGM/ICM bellow $z=1$. It also might give hints
that the high metallicity of clusters could be correlated to the dust
enrichment of the IGM/ICM. Indeed, the amount of gaseous iron in
galaxies such as the Milky Way is $\sim 0.01 $Z$_\odot$. An early
enrichment of dust in the IGM and/or the ICM, which once sputtered
will provide metals, could explain part of the iron abundances found
in the ICM at low redshifts. This hypothesis seems to be consolidated
by the few works that have investigated dust as a source for metals in
the material stripped from galaxies via dynamical removal within
already formed clusters \citep{aguirre01} or via an early IGM
enrichment at high redshift during the peak of star formation around
$z=3$ \citep{bianchi05}. The latter work stressed that only big grains
($a>0.1\mu$m) can be transported on a few 100kpc physical scale,
however leading to a very inhomogenous spatial enrichment in metals
once the dust grains are sputtered. For all these
reasons, by underestimating the metallicity at high redshifts, we
might have underestimated the amount of dust injected in the ICM at
high redshift, and thus the efficiency of dust cooling when integrated
from early epoch down to redshift zero


\section{Conclusion}

In this work, we have presented the first simulations of structure
formation investigating the effect of dust cooling on the properties
of the intra-cluster medium. We have compared simulations with
radiative $plus$ dust cooling with respect to a purely
radiative cooling simulation. We have shown that:
\begin{itemize}
\item The cooling due to dust is effective at the cluster regime and
  has a significant effect on the ``baryon driven'' statistical
  properties of cluster such as $L_{\rm X}-M$, $Y- M$,$S-M$ scaling
  relations. As an added non-gravitational cooling process dust
  changes the normalisation of these laws by a factor up to 27\% for
  the $L_{\rm X}-M$ relation, and up to 10\% for the $Y-M$ and $S-M$
  relations. On the contrary, dust has almost no effect on a ``dark
  matter driven'' scaling relation such as the $T_{\rm mw}-M$
  relation.

\item The inclusion of cooling by dust does not change significantly
the slopes of the cluster scaling laws investigated in this
paper. They compare with the results obtained in the radiative cooling
simulation model.

\item Through the implementation of our different dust models, we have
  demonstrated that the dust cooling effect at the scale of clusters
  depends strongly on the dust abundance in the ICM, but also
  on a similar proportion on the size distribution of dust
  grains. Therefore the dust efficiency is strongly dependent on the
  nature of the stripped and ejected galactic material, as well as the
  history of these injection and destruction processes along the
  cluster history. Indeed the early enrichment of dust might provide
  an already modified thermodynamical setup for the
  ``to-be-accreted'' gas at lower redshifts.

\end{itemize}

The setup of our simulations and the limitation of our dust
implementation can be considered at a ``zero order'' test with which
we demonstrated the active effect of dust on structure formation and
especially at the cluster scale. In order to go one step further, a
perspective of this work will be needed to couple the radiative
cooling function of dust with a physical and dynamical implementation
of the creation and destruction processes of dust in the IGM/ICM.

\section*{Acknowledgments}

We are deeply indebted to Peter Thomas, Orrarujee Muanwong and
collaborators for the their part in writing the original Sussex
cluster extraction software used in this work, and to Nabila Aghanim
for discussions and providing us access to the IAS (Orsay) computing
facilities where simulations were run. We thank Mauro Roncarelli,
Alain Blanchard, Peter Thomas and Nabila Aghanim for fruitful
discussions and comments on the manuscript. EP and LM acknowledge the
support of grant ANR-06-JCJC-0141. JL benefited the support of the
French-Portugese Luso actions (PAUILF -- PI: Alain Blanchard \& Pedro
Viana). AdS acknowledges support from Funda\c c\~ao Ci\^encia e
Tecnologia (FCT) under the contracts SFRH/BPD/20583/2004 and CI\^ENCIA
2007.

\begin{table*}
\begin{center}
\caption{\label{tab_best_fit} Best fir values of the parameters
  $\alpha$, $\log A$ and $\beta$ as well as their respective $1\sigma$
  errors. These values are valid within the redshift range $0<z<1$.}
\begin{tabular}{l|rrrrrr}
\hline
     & Model C & Model D5 & Model D4 & Model D3 & Model D2 & Model D1  \\
\\
\hline
$T_{\rm mw}-M$ & & & & & & \\
~~~~$\alpha_{\rm TM}$  & $0.61 \pm 0.02$ & $0.61 \pm 0.02$ & $0.61 \pm 0.02$   & $0.62 \pm 0.02$  &$0.63 \pm 0.02$  & $0.63 \pm 0.02$\\
~~~~$\log A_{\rm TM}$ & $0.195 \pm 0.002$ & $0.195 \pm 0.003$ & $0.196 \pm 0.002$    & $0.197 \pm 0.003$  &$0.201 \pm 0.002$  & $0.204 \pm 0.002$\\
~~~~$\beta_{\rm TM}$ & $-0.14 \pm 0.01$ & $-0.14 \pm 0.01$ & $-0.14 \pm 0.01$ & $-0.15 \pm 0.01$ &$-0.16 \pm 0.01$ & $-0.16 \pm 0.01$ \\
\\
\hline
$S-M$  & & & & & & \\
~~~~$\alpha_{\rm SM}$ & $0.55 \pm 0.03$    & $0.54  \pm 0.03$   & $0.54 \pm 0.03$    &$0.56 \pm 0.03$  &$0.55 \pm 0.02$  &$0.54 \pm 0.02$  \\
~~~~$\log A_{\rm SM}$ & $2.443 \pm 0.002$  & $2.444  \pm 0.002$ & $2.445 \pm 0.002$  &$2.451 \pm 0.002$  &$2.468 \pm 0.002$  &$2.488 \pm 0.02$\\
~~~~$\beta_{\rm SM}$ & $-0.33 \pm 0.01$   & $-0.34 \pm 0.01$   & $-0.34 \pm 0.01$   &$-0.36 \pm 0.01$ &$-0.40 \pm 0.01$ &$-0.42 \pm 0.01$ \\
\\
\hline
$Y-M$  & & & & & & \\
~~~~$\alpha_{\rm YM}$ & $1.74 \pm 0.03$  & $1.72 \pm 0.03$  & $1.73 \pm 0.03$  &$1.72 \pm 0.02$ &$1.74 \pm 0.02$    & $1.76 \pm 0.02$ \\
~~~~$\log A_{\rm YM}$ & $-5.909 \pm 0.002$ & $-5.907 \pm 0.002$ & $-5.910 \pm 0.002$ &$-5.914 \pm 0.002$ &$-5.933 \pm 0.002$ & $-5.957 \pm 0.002$\\
~~~~$\beta_{\rm YM}$  & $0.12 \pm 0.01$  & $0.11 \pm 0.01$  & $0.13 \pm 0.02$  &$0.13 \pm 0.01$  &$0.17 \pm 0.01$    & $0.21 \pm 0.01$\\
\\
\hline
$L_{\rm X}-M$ & & & & & & \\ 
~~~~$\alpha_{\rm LM}$ & $1.69 \pm 0.07$  & $1.68 \pm 0.07$  & $1.65 \pm 0.07$  &$1.61 \pm 0.08$  & $1.67 \pm 0.05$ & $1.67 \pm 0.05$\\
~~~~$\log A_{\rm LM}$ & $3.330 \pm 0.006$  & $3.334 \pm 0.006$  & $3.333 \pm 0.005$ &$3.323 \pm 0.005$  & $3.265 \pm 0.005$ &$3.207 \pm 0.004$ \\
~~~~$\beta_{\rm LM}$  & $0.01 \pm 0.03$ & $-0.02 \pm 0.03$ & $-0.02 \pm 0.03$ &$0.02 \pm 0.03$ &$0.18 \pm 0.03$ &$0.23 \pm 0.03$ \\
\\
\hline
\end{tabular}
\end{center}
\end{table*}

\bsp

\label{lastpage}

\end{document}